\documentclass [12 pt]{article}
\def\be{\begin{equation}}
\def\eea{\end{eqnarray}}
\def\bea{\begin{eqnarray}}
\def\ee{\end{equation}}

\usepackage[psamsfonts]{amssymb}
\usepackage{amsmath}

\author{M. Amooshahi$^{1}$ \footnote{amooshahi@sci.ui.ac.ir}
\\ $^{1}$ {\small Faculty of science, University of Isfahan ,Hezar Jarib Ave.,
Isfahan,Iran}}
\title{Canonical quantization of the electromagnetic field  in  an
anisotropic polarizable and magnetizable medium}
\begin{document}
\maketitle
\begin{abstract}
 A fully canonical
quantization of electromagnetic field  is introduced in the presence
of an anisotropic polarizable and magnetizable  medium . Two tensor
fields which couple the electromagnetic field with the medium and
have an important role in this quantization method are introduced.
The electric and magnetic polarization fields of the medium
naturally are concluded in terms of the coupling tensors and the
dynamical variables modeling the magnetodielectric medium. In
Heisenberg picture, the constitutive equations of the medium
together with the Maxwell laws are obtained as the equations of
motion of the total system and the susceptibility tensors of the
medium are calculated in terms of the coupling tensors. Following a
perturbation method the Green function related to the total system
is found and the time dependence of
electromagnetic field operators is derived.
\end{abstract}
\newpage
\section{introduction}
One of the most important quantum  dissipative  systems  is the
quantized electromagnetic field in the presence of  an absorbing
polarizable medium. In this case  there are  mainly two quantization
approaches, the phenomenological method \cite{1}-\cite{7} and the
damped polarization model\cite{8,9}. The phenomenological scheme has
been formulated on the basis of the fluctuation- dissipation
 theorem \cite{9.1}. In this method by adding a fluctuating noise
term, that is the noise polarization field, to the classical
constitutive equation of the medium, this equation is taken as the
definition of the electric polarization operator. Combination of the
Maxwell equations and the constitutive equation in the frequency
domain, gives the electromagnetic field operators in terms of the
noise polarization field and the classical Green tensor.  A set of
bosonic operators is associated with the noise polarization which
their commutation relations are given in agreement with the
fluctuation- dissipation theorem. This quantization scheme has been
quite successful in describing some electromagnetic phenomena in the
presence of a lossy dielectric medium \cite{10}-\cite{14}. The
phenomenological approach has been extended to a lossy magnetic or
anisotropic medium \cite{15}-\cite{17}. This formalism  has also
been generalized to an arbitrary linearly responding medium based on
a spatially nonlocal
conductivity tensor \cite{18}.\\
 The damped polarization model to quantize
electromagnetic field in a dispersive dielectric medium \cite{8,9}
is a canonical quantization in which the electric polarization field
of the medium applies   in the Lagrangian of the total system as a
part of the degrees of freedom of the medium. The other parts of the
degrees of freedom of the absorbing medium are related to the
dynamical variables of a heat bath describing the absorptivity
feature of the medium.  In this method the dielectric function of
the medium is found in terms of the coupling function of the heat
bath and the polarization field, so that it  satisfies  the
Kramers-Kronig relations \cite{18.1}. This quantization method has been generalized to an inhomogeneous medium\cite{19}.\\
In the present work we generalize  our previous model\cite{20,21} to
an anisotropic   dispersive magnetodielectric medium using a
canonical approach. In this formalism  the medium is modeled with
two independent collections of vector fields.  These collections
solely constitute the degrees of freedom of the medium and it is not
needed   the electric and magnetic polarization fields to be
included  in the Lagrangian of the total system as a part of the
degrees of freedom of the medium as in the Huttner-Barnett model
\cite{8}. In fact the dynamical fields modeling the dispersive
medium are able to describe both polarizability and absorptivity
features of the medium.\\
This paper is organized as follows. In Sec. 2, a Lagrangian for the
total system is proposed and  classical electrodynamics in the
presence of an anisotropic polarizable and magnetizable medium with
spatial-temporal dispersion briefly is discussed . In Sec. 3,
applying the Lagrangian introduced in Sec. 2  a fully canonical
quantization of both electromagnetic field and the dynamical
variable modeling the responding medium is demonstrated. Then in
Sec. 4,
  the constitutive equations of the  medium are obtained as the
consequences of the Heisenberg equations of the total system and
 the electric and magnetic susceptibility tensors of the
medium are calculated in terms of the parameters applied in the
theory . In Sec.5, it is shown that the Green function of the total
system in reciprocal space satisfies  an algebraic equation and a
perturbation method to obtain the Green function is introduced.
Finally in section 6 the model is modified for  media  which the
distinction between polarization and magnetization is not possible.
This paper is closed with a summary and some concluding remarks in
Sec.7 .\\

\section {Classical electrodynamics in an anisotropic magnetodielectric medium }
In order to present  a fully canonical quantization of
electromagnetic field   in the presence of an anisotropic
polarizable and magnetizable medium, we model the medium by two
independent reservoirs. Each reservoir contains a continium  of
three dimensional harmonic oscillators labeled with a continuous
parameter $\omega $. We call these two continuous sets  of
oscillators "$E$ field " and "$M$ field". The $E$ field and $M$
field describe polarizability and magnetizability of the medium,
respectively. This means that, in this approach it is not needed the
electric and magnetic polarization fields of the medium to be
appeared explicitly in the Lagrangian of the total system as a part
of the degrees of freedom of the medium , but the contribution of
the medium in the Lagrangian of the total system is related only to
the Lagrangian of the $E$ and $M$ fields and these fields completely
describe the degrees of freedom of the medium. The presence of the
"$E$ field" in the total Lagrangian is sufficient for a complete
description of both polarizability and the absorption of the medium
due to its electrically dispersive property. Also  the "$M$ field"
solely is sufficient in order to description of both magnetizability
and the absorption of the medium due to its magnetically dispersive
property. Therefore, in order to have a classical treatment of
electrodynamics in a magnetodielectric medium, we start with a
Lagrangian for the total system (medium + electromagnetic field )
which is the sum of three parts
\begin{equation}\label{c1}
 L(t)= L_{res}+ L_{em}+L_{int}
 \end{equation}
where $L_{res}$ is the part related to the degrees of freedom of
the medium and is the sum of the Lagrangians of the $E$ and $M$
fields
\begin{equation}\label{c2}
 L_{res}= L_{e}+L_{m}
 \end{equation}
 where
 \begin{equation}\label{c2.1}
L_{e}=\int _0^\infty d\omega \int d^3r \left[ \frac{1}{2}
 \dot{\vec{X}}_\omega \cdot  \dot{\vec{X}}_\omega
 -\frac{1}{2}\omega^2
 \vec{X}_\omega \cdot  \vec{X}_\omega\right]
\end{equation}
 and
\begin{equation}\label{c2.2}
L_{m}=\int _0^\infty d\omega \int d^3r \left[ \frac{1}{2}
 \dot{\vec{Y}}_\omega \cdot  \dot{\vec{Y}}_\omega
 -\frac{1}{2}\omega^2
 \vec{Y}_\omega \cdot  \vec{Y}_\omega\right]
 \end{equation}
 Here the fields $ \vec{X}_\omega $ and $\vec{Y}_\omega $  are the dynamical variables of
 the $E$ and $M$ fields, respectively .

  In (\ref{c1}) $L_{em}$ is the contribution of the electromagnetic
 field in the Lagrangian of the total system
\begin{equation}\label{c3}
L_{em}= \int d^3r \left[ \frac{1}{2}\varepsilon _0
\vec{E}^2-\frac{\vec{B}^2}{2\mu_0}\right]
\end{equation}
and $L_{int}$ is the part describing the interaction of the
electromagnetic field with the  medium
\begin{eqnarray}\label{c4}
&&L_{int}= \int _0^\infty d \omega \int d^3 r \int d^3 r' f_{ij}(
\omega , \vec{r} , \vec{r'}) E^i(\vec{r},t) \vec{X}_\omega
^j(\vec{r'} ,t)+\nonumber\\
&&\int _0^\infty d \omega \int d^3 r \int d^3 r' g_{ij}( \omega ,
\vec{r} , \vec{r'}) B^i(\vec{r},t) \vec{Y}_\omega ^j(\vec{r'} ,t)
\end{eqnarray}
The contributions $L_{e}$ and $L_m$ in the Lagrangian $L_{res}$  are
equivalent to the consequences of
  diagonalization processes   of the matter fields in the Huttner-Barnet model [8]. That is,  $L_{e}$  ( $L_{m}$ )
 is equivalent to the  diagonalization of the contributions of related to three parts in the Huttner-Barnet model:
  the dynamical variable describing the electric polarization (magnetic polarization)of the medium ,
 a heat-bath $B$ ( $B'$) interacting with the electric polarization (magnetic polarization )  and the interaction term
  between the heat-bath $B$ ( $B'$) and the electric polarization (magnetic polarization). In the present approach modeling the medium , in a phenomenological
way,  with two independent set of oscillators the lengthy
diagonalization processes have been eliminated in the start of this
quantization scheme. Particularly the diagonalization
processes may be more tremendous for an anisotropic medium.\\

In equations (\ref{c3}) and (\ref{c4}) $\vec{E}=-
\frac{\partial\vec{A}}{\partial t}-\vec{\nabla\varphi}$ and
$\vec{B}=\nabla\times\vec{A}$  are electric and magnetic fields
respectively, where  $\vec{A}$ and  $\varphi $ are the vector and
the scalar potentials. The tensors $ f $ and $g$ in (\ref{c4}) are
called the coupling tensors of the medium with electromagnetic field
and for an inhomogeneous  medium  are dependent on the both position
vectors $ \vec{r}$ and $\vec{r'}$. The coupling tensors are the key
parameters of this theory. As was mentioned above, it is not
 needed    the electric and magnetic polarization fields of the
medium explicitly to be appeared in the total Lagrangian
(\ref{c1})-(\ref{c4}) as a part of degrees of freedom of the medium
. As we will see, the electric polarization( magnetic polarization )
of the medium is obtained  in terms of the coupling tensor $f$
 ( $g$ ) and the dynamical variables $ \vec{X}_\omega $ \ ( $
\vec{Y}_\omega $). Also the electric susceptibility tensor (
magnetic susceptibility tensor) of the medium naturally  will
expressed in terms of the coupling tensor  $f$ ( $g$ ). The coupling
tensors $f$ and $g$ are appeared as common   factors in both the
noise polarization fields and the  susceptibility tensors of the
medium, so that  for the free space the susceptibility tensors
together with the noise polarizations become identically  zero and
this quantization scheme is reduced to the usual quantization of
electromagnetic field  in free space. Furthermore  when the medium
tends   to a non-absorbing one, the coupling tensors and the noise
polarizations  tend also to zero and this quantization
method is reduced to the quantization in a non-absorbing medium \cite{20,22}.\\

In order to prevent some difficulties  with a non-local Lagrangian
such as in (\ref{c4}), it is the easiest way to work in the
reciprocal space and write all the fields and the coupling tensors
$f\ ,\ g$ in terms of their spatial Fourier transforms. For example
the dynamical variable $ \vec{X}_\omega $ can be written as
\begin{equation}\label{c5}
\vec{X}_\omega(\vec{r},t)=\frac{1}{\sqrt{(2\pi)^3)}}\int d^3k\
\vec{\underline{X}}_\omega(\vec{k},t)\ e^{\imath \vec{k}\cdot
\vec{r}}
\end{equation}
Since we are concerned with real valued fields in the total
Lagrangian (\ref{c1})-(\ref{c4}), we have
$\vec{\underline{X}}_\omega^*(\vec{k},t)=\vec{\underline{X}}_\omega(-\vec{k},t)$
for the   field $ \vec{X}_\omega(\vec{r},t)$ and the other dynamical
fields in this Lagrangian. Similarly the real valued coupling
tensors $f$ and $g$ can be expressed in reciprocal space as
\begin{eqnarray}\label{c6}
f_{ij}(\omega,\vec{r},\vec{r'})&=&\frac{1}{(2\pi)^3} \int d^3k
\int d^3k'  \underline{f}_{ij}(\omega , \vec{k} , \vec{k'})
e^{\imath \vec{k}\cdot \vec{r}- \imath \vec{k'}\cdot
\vec{r'}}\nonumber\\
g_{ij}(\omega,\vec{r},\vec{r'})&=&\frac{1}{(2\pi)^3} \int d^3k
\int d^3k'  \underline{g}_{ij}(\omega , \vec{k} , \vec{k'})
e^{\imath \vec{k}\cdot \vec{r}- \imath \vec{k'}\cdot \vec{r'}}
\end{eqnarray}
which obey the following conditions
\begin{eqnarray}\label{c6.1}
\underline{f}_{ij}(\omega , \vec{k} ,
\vec{k'})&=&\underline{f}_{ij}^*(\omega , -\vec{k} ,
-\vec{k'})\nonumber\\
\underline{g}_{ij}(\omega , \vec{k} ,
\vec{k'})&=&\underline{g}_{ij}^*(\omega , -\vec{k} , -\vec{k'})
\end{eqnarray}
 The number of independent variables can be recovered by restricting the integrations  to the half space $k_z\geq 0$. The total Lagrangian
(\ref{c1})-(\ref{c4}) is then obtained as
\begin{equation}\label{c7}
 \underline{L}(t)= \underline{L}_{res}(t)+
\underline{L}_{em}(t)+\underline{L}_{int}(t)
\end{equation}
\begin{equation}\label{c8}
\underline{L}_{res}(t)=\int_0^\infty d\omega \int'd^3k
\left(|\dot{\underline{\vec{X}}}_\omega|^2-\omega^2|\underline{\vec{X}}_\omega|^2\right)
 +\int_0^\infty d\omega\int'd^3k
\left(|\dot{\underline{\vec{Y}}}_\omega|^2-\omega^2|\underline{\vec{Y}}_\omega|^2\right)
\end{equation}
\begin{equation}\label{c9}
\underline{L}_{em}(t)= \int'd^3k
\left(\varepsilon_0|\dot{\underline{\vec{A}}}|^2 +\varepsilon_0
|\vec{k}\
\underline{\varphi}|^2-\frac{|\vec{k}\times\underline{\vec{A}}|^2}{\mu_0}
\right) +\varepsilon_0\int' d^3k\left( -\imath \vec{k}\cdot
\dot{\underline{\vec{A}}}\ \  \underline{\varphi}^*+h.c\right)
\end{equation}
\begin{eqnarray}\label{c10}
&&\underline{L}_{int}(t)=\nonumber\\
&-&\int_0^\infty d\omega\ \int' d^3q \int' d^3p\left[ \left(
\underline{\dot{\vec{A}}}(\vec{q},t)+\imath\vec{ q}\ \
\underline{\varphi}(\vec{q},t)\right)\cdot\underline{f}( \omega,
-\vec{q} ,\vec{p})
\cdot\underline{\vec{X}}_\omega(\vec{p},t)+h.c\right]\nonumber\\
&-&\int_0^\infty d\omega\ \int' d^3q \int' d^3p\left[ \left(
\underline{\dot{\vec{A^*}}}(\vec{q},t)-\imath\vec{ q}\
\underline{\varphi}^*(\vec{q},t)\right)\cdot\underline{f}( \omega,
\vec{q} ,\vec{p})\cdot
\underline{\vec{X}}_\omega(\vec{p},t)+h.c\right]\nonumber\\
&+&\int_0^\infty d\omega\ \int' d^3q \int' d^3p\left[ \left(
\imath\vec{q}\times
\underline{\vec{A}}(\vec{q},t)\right)\cdot\underline{g}( \omega,
-\vec{q} ,\vec{p})
\cdot\underline{\vec{Y}}_\omega(\vec{p},t)+h.c\right]\nonumber\\
&+&\int_0^\infty d\omega\ \int' d^3q \int' d^3p\left[\left(
-\imath\vec{q}\times \underline{\vec{A^*}}(\vec{q},t)\right)\cdot
\underline{g}( \omega, \vec{q} ,\vec{p})\cdot
\underline{\vec{Y}}_\omega(\vec{p},t)+h.c\right]
\end{eqnarray}
where $\displaystyle \int'd^3k $ implies the integration  over the
half space $ k_z\geq 0$ ( hereafter , we apply the symbol
$\displaystyle \int'd^3k $ for the integration on the half space $
k_z\geq 0 $ and $\displaystyle \int d^3k $ for the integration on
the total  reciprocal space). In the reciprocal space the total
Lagrangian (\ref{c7})- (\ref{c10}) do not involve the space
derivatives of the dynamical variables of the system and
 the classical equations of the motion of the
system can be obtained  using  the principle of the Hamilton's least
action, $ \displaystyle \delta \int dt \ \underline{L}(t)=0$. These
equations are the Euler-Lagrange equations. For the vector potential
$ \underline{\vec{A}}(\vec{k},t)$ and the scalar potential $
\underline{\varphi}(\vec{k},t)$ we find
\begin{eqnarray}\label{c11}
&& \frac{d}{d t}\left(\frac{\delta\underline{L}}{\delta
\left(\underline{\dot{A}}^*_i(\vec{k},t)\right)}\right)-\frac{\delta\underline{L}}{\delta\left(\underline{A}^*_i(\vec{k},t)\right)}=0
\hspace{2.00 cm} i=1,2,3\nonumber\\
\Longrightarrow&&\mu_0\varepsilon_0\underline{\ddot{\vec{A}}}(\vec{k},t)+\mu_0\varepsilon_0
\ \imath\vec{k}\ \underline{\dot{\varphi}}(\vec{k},t)-\vec{k}\times\left(\vec{k}\times\underline{\vec{A}}(\vec{k},t)\right)=\nonumber\\
&&\mu_0\ \underline{\dot{\vec{P}}}(\vec{k},t)+\imath\mu_0\
\vec{k}\times\underline{\vec{M}}(\vec{k},t)
\end{eqnarray}
\begin{eqnarray}\label{c12}
&&\frac{d}{d t}\left(\frac{\delta\underline{L}}{\delta
\left(\underline{\dot{\varphi}}^*(\vec{k},t)\right)}\right)-\frac{\delta\underline{L}}{\delta\left(\underline{\varphi}^*(\vec{k},t)\right)}=0\nonumber\\
\Longrightarrow&&-\varepsilon_0\
\imath\vec{k}\cdot\underline{\dot{\vec{A}}}(\vec{k},t)+\varepsilon_0|\vec{k}|^2\underline{\varphi}(\vec{k},t)
=-\imath\vec{k}\cdot\underline{\vec{P}}(\vec{k},t)
\end{eqnarray}
 for any wave vector $\vec{k}$ in
the half space $k_z\geq 0$ where
\begin{equation}\label{c13}
\underline{\vec{P}}(\vec{k},t)=\int_0^\infty d\omega\int d^3p \
\underline{f}( \omega , \vec{k},\vec{p})\cdot\
\underline{X}_\omega(\vec{p},t)
\end{equation}
\begin{equation}\label{c13.1}
\underline{\vec{M}}(\vec{k},t)=\int_0^\infty d\omega\int d^3p \
\underline{g}( \omega , \vec{k},\vec{p})\cdot\
\underline{Y}_\omega(\vec{p},t)
\end{equation}
are respectively the spatial Fourier  transforms of the electric and
magnetic polarization densities of the medium and it has been  used
from the relations (\ref{c6.1}). Therefore in this method the
polarization fields of the medium  are naturally  concluded in terms
of the coupling tensors $f , g$ and the dynamical variables of the
$E$ and $M$ fields.  Similarly the Euler-Lagrange equations for the
fields $ \underline{\vec{X}}_\omega $ and $
\underline{\vec{Y}}_\omega $ for any vector $\vec{k}$ in the half
space $k_z\geq 0$  are easily obtained as
\begin{eqnarray}\label{c14}
&& \frac{d}{d t}\left(\frac{\delta\underline{L}}{\delta
\left(\underline{\dot{X}}^*_{\omega
i}(\vec{k},t)\right)}\right)-\frac{\delta\underline{L}}{\delta\left(\underline{X}^*_{\omega
i}(\vec{k},t)\right)}=0
\hspace{2.00 cm} i=1,2,3\nonumber\\
\Longrightarrow &&\underline{\ddot{\vec{X}}}_\omega
(\vec{k},t)+\omega^2\ \underline{\vec{X}}_\omega(\vec{k},t)= -\int
d^3q\  \underline{f}^\dag( \omega , \vec{q}, \vec{k})\cdot
\left(\underline{\dot{\vec{A}}}(\vec{q},t)+\imath\vec{q}\underline{\varphi}(\vec{q},t)\right)\nonumber\\
&&
\end{eqnarray}
\begin{eqnarray}\label{c15}
&& \frac{d}{d t}\left(\frac{\delta\underline{L}}{\delta
\left(\underline{\dot{Y}}^*_{\omega
i}(\vec{k},t)\right)}\right)-\frac{\delta\underline{L}}{\delta\left(\underline{Y}^*_{\omega
i}(\vec{k},t)\right)}=0
\hspace{2.00 cm} i=1,2,3\nonumber\\
\Longrightarrow &&\underline{\ddot{\vec{Y}}}_\omega
(\vec{k},t)+\omega^2\ \underline{\vec{Y}}_\omega(\vec{k},t)=\int
d^3q \ \underline{g}^\dag( \omega , \vec{q}, \vec{k})\cdot
\left(\imath\ \vec{q}\times\underline{\vec{A}}(\vec{q},t)\right)\nonumber\\
&&
\end{eqnarray}
where  $\underline{f}^\dag and  \underline{g}^\dag$  are the
hermitian conjugate of the tensors $ \underline{f}$ and
$\underline{g}$, respectively.
\section{ Canonical quantization}
Following the standard approach, we choose the Coulomb gauge
$\vec{k}\cdot\underline{\vec{A}}(\vec{k},t)=0$ to quantize
electromagnetic field. In this gauge the vector potential $\vec{A}$
is a purely transverse field and can be decomposed along the unit
polarization vectors $\vec{e}_{\lambda\vec{k}}\hspace {1.00
cm}\lambda=1,2$  which are orthogonal to each other and to the wave
vector $\vec{k}$.
\begin{equation}\label{c16}
\underline{\vec{A}}(\vec{k},t)=\sum_{ \lambda=1}^2\
\underline{A}_\lambda( \vec{k},t)\vec{e}_{\lambda\vec{k}}
\end{equation}
 Although the vector potential is purely transverse, but the dynamical fields
  $\underline{\vec{X}}_\omega$ and $\underline{\vec{Y}}_\omega$  may  have both transverse and
  longitudinal parts and can be expanded along the three mutually   orthogonal unit vectors
  $\vec{e}_{\lambda\vec{k}}\hspace {0.50 cm}\lambda=1,2$ and
  $\vec{e}_{3\vec{k}}=\hat{k}=\frac{\vec{k}}{|\vec{k}|}$ as
  \begin{eqnarray}\label{c17}
\underline{\vec{X}}_\omega(\vec{k},t)&=&\sum_{ \lambda=1}^3
\underline{X}_{\omega\lambda}( \vec{k},t)\vec{e}_{\lambda\vec{k}}\nonumber\\
\underline{\vec{Y}}_\omega(\vec{k},t)&=&\sum_{ \lambda=1}^3
\underline{Y}_{\omega\lambda}( \vec{k},t)\vec{e}_{\lambda\vec{k}}
\end{eqnarray}
Furthermore in Coulomb gauge  the Euler- Lagrange equation
(\ref{c12}) can be used to eliminate  the extra degree of freedom
$\underline{\varphi}(\vec{k},t)$ from the Larangian of the system
\begin{equation}\label{c18}
\underline{\varphi}(\vec{k},t)=-\frac{\imath\vec{k}\cdot\underline{\vec{P}}(\vec{k},t)}{\varepsilon_0
|\vec{k}|^2 }
\end{equation}
The total Lagrangian  (\ref{c7})-(\ref{c10}) can now be rewritten
in terms of the independent  dynamical variables
$\underline{A}_\lambda\hspace{0.50 cm}\lambda=1,2$ and $
\underline{\vec{X}}_{\omega\lambda} \ ,\
\underline{\vec{Y}}_{\omega\lambda}  \hspace{0.50
cm}\lambda=1,2,3 $ which constitute completely the coordinates of
the total system
\begin{eqnarray}\label{c19}
 \underline{L}(t)&=&\int_0^\infty d\omega\ \int'd^3k
\sum_{\lambda=1}^3\left(|\dot{\underline{\vec{X}}}_{\omega\lambda}|^2-\omega^2|\underline{\vec{X}}_{\omega\lambda}|^2
 +|\dot{\underline{\vec{Y}}}_{\omega\lambda}|^2-\omega^2|\underline{\vec{Y}}_{\omega\lambda}|^2\right)\nonumber\\
& +& \int'd^3k
\sum_{\lambda=1}^2\left(\varepsilon_0|\dot{\underline{{A}}_\lambda}|^2
-\frac{|\vec{k}\ \underline{A}_{\lambda}|^2}{\mu_0} \right)
-\frac{1}{\varepsilon_0} \int' d^3k\
\frac{|\imath\vec{k}\cdot\underline{\vec{P}}|^2}{|\vec{k}|^2}\nonumber\\
& +&\int' d^3k \left[\left(-
\sum_{\lambda=1}^2\underline{\dot{A}}_\lambda
\vec{e}_{\lambda\vec{k}}\right)\cdot \underline{\vec{P}}^*+ \left(
\imath\vec{k}\times\sum_{\lambda=1}^2\underline{A}_\lambda
\vec{e}_{\lambda\vec{k}}\right)\cdot
\underline{\vec{M}}^*+h.c\right]\nonumber\\
&&
\end{eqnarray}
where the polarizations $\underline{\vec{P}}$ and $
\underline{\vec{M}}$ have been  defined previously in terms of the
dynamical variables of the $E$ and $M$ fields in equations
(\ref{c13}) and (\ref{c13.1}). The Lagrangian (\ref{c19}) can now be
used to define the canonical conjugate momenta  of the system. For
any wave vector $\vec{k}$ in half space $k_z\geq 0$ this momenta are
defined as
\begin{eqnarray}\label{c23}
-\underline{D}_\lambda (\vec{k},t)&=&\frac{\delta
\underline{L}}{\delta \left(\underline{\dot{A}}^*_\lambda
(\vec{k},t)\right)}=
\varepsilon_0\dot{A}_\lambda (\vec{k},t)-\vec{e}_{\lambda\vec{k}}\cdot\underline{\vec{P}}(\vec{k},t)\nonumber\\
\underline{Q}_{\omega\lambda}(\vec{k},t)&=&\frac{\delta
\underline{L}}{\delta\left(
\underline{\dot{X}}^*_{\omega\lambda}(\vec{k},t)\right)}=
\underline{\dot{X}}_{\omega\lambda}(\vec{k},t)\nonumber\\
\underline{\Pi}_{\omega\lambda}(\vec{k},t)&=&\frac{\delta
\underline{L}}{\delta
\left(\underline{\dot{Y}}^*_{\omega\lambda}(\vec{k},t)\right)}=
\underline{\dot{Y}}_{\omega\lambda}(\vec{k},t)
\end{eqnarray}
The total system are quantized canonically in a standard method by
imposing equal-time commutation relations between the coordinates of
the system and their conjugates variables as follows
\begin{eqnarray}\label{c24}
\left[ \underline{A}^*_\lambda ( \vec{k},t) \ ,\
-\underline{D}_{\lambda'}( \vec{k'} ,
t)\right]&=&\imath\hbar\delta_{\lambda\lambda'}\delta(\vec{k}-\vec{k'})\nonumber\\
\left[ \underline{X}^*_{\omega\lambda }( \vec{k},t) \ ,\
\underline{Q}_{\omega'\lambda'}( \vec{k'} ,
t)\right]&=&\imath\hbar\delta_{\lambda\lambda'}\delta(\omega-\omega')\delta(\vec{k}-\vec{k'})\nonumber\\
\left[ \underline{Y}^*_{\omega\lambda }( \vec{k},t) \ ,\
\underline{\Pi}_{\omega'\lambda'}( \vec{k'} ,
t)\right]&=&\imath\hbar\delta_{\lambda\lambda'}\delta(\omega-\omega')\delta(\vec{k}-\vec{k'})\nonumber\\
&&
\end{eqnarray}
Using the Lagrangian (\ref{c19}) and the conjugates momenta
introduced in (\ref{c23}) the Hamiltonian of the total system can be
written in the form
\begin{eqnarray}\label{c25}
H(t)&=&\int' d^3k \sum_{\lambda=1}^2\left(
\frac{|\underline{D}_{\lambda}-\vec{e}_{\lambda\vec{k}}\cdot
\underline{\vec{P}}|^2}{\varepsilon_0}+\frac{|\vec{k}\underline{A}_\lambda
|^2}{\mu_0}\right)+\int' d^3k\frac{|\imath\vec{k}\cdot\underline{\vec{P}}|^2}{\varepsilon_0|\vec{k}|^2}\nonumber\\
&-&\int' d^3k \left[\left(
\imath\vec{k}\times\sum_{\lambda=1}^2\underline{A}_{\lambda}\vec{e}_{\lambda\vec{k}}\right)\cdot
\underline{\vec{M}}^*+h.c\right]
\nonumber\\
&+&\int_0^\infty d\omega\int' d^3k\sum_{\lambda=1}^3\left( |\
\underline{Q}_{\omega \lambda}|^2+\omega^2|\ \underline{X}_{\omega
\lambda}|^2 + |\ \underline{\Pi}_{\omega \lambda}|^2+\omega^2|\
\underline{Y}_{\omega\lambda}|^2\right)\nonumber\\
&&
\end{eqnarray}
\subsection{Maxwell equations}
Using Heisenberg equations for the operators $\underline{D}_\lambda
$ and $\underline{A}_\lambda $ , that is
\begin{eqnarray}\label{c26}
&&\underline{\dot{A}}_\lambda (
\vec{k},t)=\frac{\imath}{\hbar}\left[ H \ , \underline{A}_\lambda (
\vec{k},t)\right]= -\ \frac{\underline{D}_\lambda (
\vec{k},t)-\vec{e}_{\lambda\vec{k}}\ \cdot\ \underline{\vec{P}}(\vec{k},t)}{\varepsilon_0}\nonumber\\
&&\underline{\dot{D}}_\lambda (
\vec{k},t)=\frac{\imath}{\hbar}\left[ H \ , \underline{D}_\lambda (
\vec{k},t)\right]=\frac{|\vec{k}|^2}{\mu_0}\underline{A}_\lambda (
\vec{k},t) - \vec{e}_{\lambda\vec{k}}\ \cdot\  \left(
\imath\vec{k}\times
\underline{\vec{M}}(\vec{k},t)\right)\nonumber\\
&&
\end{eqnarray}
Maxwell equations can be obtained as is expected. Multiplying  both
sides of Eqs (\ref{c26}) in the polarization unit vector
$\vec{e}_{\lambda\vec{k}}$ and then   summation on polarization
index $ \lambda=1,2 $ and using (\ref{c18})clearly give us
\begin{equation}\label{c27}
\underline{\vec{D}}(\vec{k},t)=\varepsilon_0
\underline{\vec{E}}(\vec{k},t)+\underline{\vec{P}}(\vec{k},t)
\end{equation}
\begin{equation}\label{c27.1}
\underline{\dot{\vec{D}}}(\vec{k},t)=\imath\vec{k}\times\underline{\vec{H}}
\end{equation}
where \ $\displaystyle
\underline{D}=\sum_{\lambda=1}^2\underline{\dot{D}}_\lambda
\vec{e}_{\lambda\vec{k}}$\  plays  the role of the displacement
field,
$\displaystyle\underline{\vec{E}}=-\underline{\dot{\vec{A}}}-\imath\vec{k}\underline{\varphi}=
-\underline{\dot{\vec{A}}}-\frac{\hat{k}(\hat{k}\cdot\underline{\vec{P}})}{
\varepsilon_0}$ is  the total electric field  and
$\displaystyle\underline{\vec{H}}(\vec{k},t)=\frac{\imath\vec{k}\times
\underline{\vec{A}}}{\mu_0}-\underline{\vec{M}}$ is the magnetic
induction field.\\
\section{ The constitutive equations of the medium}
It is well known that the constitutive equations of a responding
medium  are  the consequences  of the interaction of the medium with
electromagnetic field. Any quantization method such as this theory,
in which the medium enter directly in the process of quantization
and the interaction of the medium with the vacuum field is
explicitly  given, must be able to give the constitutive equations
of the medium using the Heisenberg equations of the total system.
Also the susceptibility tensors of the medium which are a measure
for the polarizability of the medium should be specified  in terms
of the parameters describing the interaction of the medium with
electromagnetic field. In this section, we find the constitutive
equations of the medium  using the Heisenberg equations of the
dynamical fields modeling the medium. As we do this, the electric
and magnetic susceptibility tensors of the medium  are naturally
found in terms of the coupling tensors $f$ and $g$. In the
Heisenberg picture, using  the commutation relations (\ref{c24}) and
the total Hamiltonian (\ref{c25}), the equations of motion for the
canonical variables $ \underline{X}_{\omega\lambda}\  ,\
\underline{Q}_{\omega\lambda} \hspace{00.50 cm} \lambda=1,2,3 $ for
each wave vector in the half space $ k_z\geq 0 $ follow as
\begin{eqnarray}\label{c28}
\underline{\dot{X}}_{\omega\lambda}(\vec{k},t)
&=&\frac{\imath}{\hbar}[ H ,
\underline{X}_{\omega\lambda}(\vec{k},t)]=\underline{Q}_{\omega\lambda}(\vec{k},t)\nonumber\\
\underline{\dot{Q}}_{\omega\lambda}(\vec{k},t)&=&[ H ,
\underline{Q}_{\omega\lambda}(\vec{k},t)]= -
\omega^2\underline{X}_{\omega\lambda}(\vec{k},t) + \int d^3q
\underline{f}^*_{ij}( \omega , \vec{q} , \vec{k})
\underline{E}^i(\vec{q},t)
 e^{j}_{\lambda\vec{k}}\nonumber\\
 &&
\end{eqnarray}
where
$\displaystyle\underline{\vec{E}}(\vec{q},t)=-\underline{\dot{\vec{A}}}(\vec{q},t)-\frac{\hat{q}\left(\hat{q}\cdot\underline{\vec{P}}(\vec{q},t)\right)}{\varepsilon_0}$
is the Fourier transform of the total electric field. Multiplying
both sides of Eqs. (\ref{c28})  in the mutually orthogonal unit
vectors $ \vec{e}_{\lambda\vec{k}}$  and then summation on
$\lambda=1,2 ,3$ and using the completeness relations $\displaystyle
\sum_{\lambda=1}^3
e^{i}_{\lambda\vec{k}}e^{j}_{\lambda\vec{k}}=\delta_{ij}$ we find
\begin{equation}\label{c29}
\underline{\ddot{\vec{X}}}(\vec{k},t)+\omega^2
\underline{\vec{X}}(\vec{k},t)= \int d^3q\ \underline{f}^\dag
(\omega , \vec{q}, \vec{k})\ \cdot\ \underline{\vec{E}}(\vec{q},t)
\end{equation}
This equation can be integrated formally as
\begin{eqnarray}\label{c30}
\underline{\vec{X}}_\omega(\vec{k},t)&=&\underline{\vec{Q}}_\omega(\vec{k},0)\frac{\sin
\omega t }{\omega}+ \underline{\vec{X}}_\omega(\vec{k},0)\cos
\omega t\nonumber\\
&+&\int _0^t d t' \frac{\sin\omega(t-t')}{\omega}\int d^3q\
\underline{f}^\dag( \omega , \vec{q} , \vec{k} )\cdot
\underline{\vec{E}}(\vec{q},t')
\end{eqnarray}
Now, by substituting $\underline{\vec{X}}_\omega(\vec{k},t)$ from
(\ref{c30}) in the definition of electric polarization density given
by (\ref{c13}), we find the constitutive equation of the medium in
reciprocal space relating the electric polarization to the electric
field
\begin{equation}\label{c31}
\underline{\vec{P}}(\vec{k},t)=\underline{\vec{P}}_N(\vec{k},t)+\varepsilon_0\int_0^{|t|}
dt' \int d^3q\ \underline{\chi}^e(\vec{k},\vec{q},
|t|-t')\cdot\underline{\vec{E}}(\vec{q},\pm t')
\end{equation}
where the upper (lower) sign corresponds to $t>0$ ($ t<0 $). Here
the $\underline{\chi}^e$ is the spatial Fourier transform of the
electric susceptibility tensor
\begin{equation}\label{c32}
\underline{\chi}^e(\vec{k},\vec{q}, t)=\left\{
\begin{array}{cc}
  \frac{1}{\varepsilon_0}\int_0^\infty d\omega \frac{\sin \omega t}{\omega}\int d^3p \ \underline{f}(\omega , \vec{k} , \vec{p}) \cdot
  \underline{f}^\dag(\omega , \vec{q} , \vec{p}) & t> 0 \\
  \\
  0 & t\leq 0
\end{array}\right.
\end{equation}
and is guaranteed to possess solutions for the coupling tensor $
\underline{f} $ in terms of the temporal Fourier transform of
$\underline{\chi}^e$ using a type of eigenvalue problem\cite{18}. It
must be pointed out that  for a given $\underline{\chi}_e$  the
coupling tensor $ \underline{f} $ satisfying the relation
(\ref{c32}) is not unique. In fact if $ \underline{f} $ satisfy
(\ref{c32}), for a given susceptibility tensor ,then
\begin{equation}\label{c33}
\underline{f}'(\omega , \vec{k},\vec{q})=\int d^3p'\
\underline{f}(\omega , \vec{k},\vec{p'})\ \cdot\
\underline{A}(\omega , \vec{q},\vec{p'})
\end{equation}
also satisfy the (\ref{c32}) where  $ \underline{A}$ is a tensor
with orthogonality condition
\begin{equation}\label{c34}
\int d^3p\ \underline{A}(\omega , \vec{p},\vec{p'})\ \cdot\
\underline{A}^\dag(\omega , \vec{p},\vec{p''})=I\delta^3(
\vec{p'}-\vec{p''})
\end{equation}
Although for a given $\underline{\chi}^e$ various choices of the
tensor $\underline{f}$ satisfying ( \ref{c32})  affect  the
space-time dependence of electromagnetic field operators, but all of
these choices are equivalent and the commutation relations between
electromagnetic field operators remain unchanged \cite{22}.  That
is, the commutation relations between electromagnetic operators and
the physical observables finally are dependent only on the given
susceptibility tensor and not on the coupling functions $\underline{f} , \underline{g}$.\\
In (\ref{c31}),  $\underline{\vec{P}}_N$ is the spatial Fourier
transform of the noise electric polarization density which is
expressed in terms of the the dynamical variables of the "$E$ field"
at $t=0$
\begin{equation}\label{c35}
\underline{\vec{P}}_N(\vec{k},t)=\int_0^\infty d\omega\int d^3p \
\underline{f}( \omega ,
\vec{k},\vec{p})\cdot\left(\underline{\vec{Q}}_\omega(\vec{p},0)\frac{\sin
\omega t }{\omega}+ \underline{\vec{X}}_\omega(\vec{p},0)\cos
\omega t\right)
\end{equation}
In a similar fashion the constitutive equation relating the magnetic
polarization density of the medium to the magnetic field
$\underline{B}(\vec{q},t)=
\imath\vec{q}\times\underline{A}(\vec{q},t)$ can be obtained
straightforwardly using  the Heisenberg equations for the conjugate
dynamical variables $ \underline{Y}_{\omega\lambda}$ and
$\underline{\Pi}_{\omega\lambda}$ as the form
\begin{equation}\label{c36}
\underline{\vec{M}}(\vec{k},t)=\underline{\vec{M}}_N(\vec{k},t)+\frac{1}{\mu_0}\int_0^{|t|}
dt' \int d^3q\ \underline{\chi}^m(\vec{k},\vec{q},
|t|-t')\cdot\underline{\vec{B}}(\vec{q},\pm t')
\end{equation}
 where
the $\underline{\chi}^m$ is the spatial Fourier transform of the
magnetic susceptibility tensor which is written in terms the
coupling tensor $\underline{g}$ as
\begin{equation}\label{c37}
\underline{\chi}^m(\vec{k},\vec{q}, t)=\left\{
\begin{array}{cc}
  \mu_0\int_0^\infty d\omega \frac{\sin \omega t}{\omega}\int d^3p \ \underline{g}(\omega , \vec{k} , \vec{p}) \cdot
  \underline{g}^\dag(\omega , \vec{q} , \vec{p}) & t> 0 \\
  \\
  0 & t\leq 0
\end{array}\right.
\end{equation}
and
 \begin{equation}\label{c38}
\underline{\vec{M}}_N(\vec{k},t)=\int_0^\infty d\omega\int d^3p \
\underline{g}( \omega ,
\vec{k},\vec{p})\cdot\left(\underline{\vec{\Pi}}_\omega(\vec{p},0)\frac{\sin
\omega t }{\omega}+ \underline{\vec{Y}}_\omega(\vec{p},0)\cos \omega
t\right)
\end{equation}
is the noise magnetic  polarization density. Therefore modeling an
anisotropic   magnetodielectric medium  with two independent
collections of harmonic oscillators, that is, the $E$ and $M$ field,
we successfully have  constructed a fully canonical quantization of
electromagnetic field in the presence of such a medium, so that the
Maxwell equations together with the constitutive equations of the
medium have been obtained from the Heisenberg equations of the total
system. In this method the susceptibility tenors of the medium are
concluded in terms of the coupling tensors which are the key
parameters of this theory and are describing the coupling of the
electromagnetic field with the medium.\\

The Equations  (\ref{c32}) and (\ref{c37}) relate the parameters of
this theory, that is  the coupling tensors $f$ and $g$, to  the
physical quantities $\underline{\chi}^e$ and  $\underline{\chi}^m$ .
If the coupling tensors $f$ and $g$ are specified, so that the right
hands  of  (\ref{c32}) and (\ref{c37}) become  identical to   the
electric and magnetic susceptibility tensors of the medium, then
according to the constitutive equations (\ref{c31}) and (\ref{c36})
 the operators $\underline{\vec{P}}$ and
$\underline{\vec{M}}$ defined by (\ref{c13}) and (\ref{c13.1}) are
respectively the polarization and magnetization of the medium.\\
 There are media for which the polarization ( the
magnetization) is dependent on both the electric and magnetic field.
This
 quantization scheme can be generalized for such media if in the
 interaction Lagrangian
  (\ref{c4}) each of the dynamical variables
 $\vec{X}_\omega$ and $\vec{Y}_\omega$ is interacted with both
 the electric and magnetic field.\\

\section{ Space -time dependence of the electromagnetic field operators}
Our goal for this section is to determine the complete expressions
of the electromagnetic field operators for both negative and
positive times. The Heisenberg equations of the total system, that
is the Maxwell equation (\ref{c27.1}) together with the constitutive
equations (\ref{c27}), (\ref{c31}) and (\ref{c36}), constitute a set
of coupled linear equations which can be solved in terms of the
initial conditions by the temporal Laplace transformation. This
technique has been used previously  in the damped polarization
model\cite{19}. For an arbitrary time-dependent  operator
$\Gamma(t)$  the backward and forward Laplace transformations,
denoted respectively, by $\Gamma^b(s)$ and $ \Gamma^f(\rho)$ are
defined as
\begin{eqnarray}\label{c39}
\Gamma^b(s)&=&\int_0^\infty dt\ e^{-s t}\
\Gamma(-t)\nonumber\\
\Gamma^f(s)&=&\int_0^\infty dt\ e^{-s t}\ \Gamma(t)
\end{eqnarray}
Carrying out the backward and forward  Laplace transformations of
Eqs.(\ref{c27}) , (\ref{c27.1}), (\ref{c31}) , (\ref{c36}) and  $
\imath\vec{k}\times \underline{\vec{E}}(\vec{k},t)= -
\underline{\dot{\vec{B}}}(\vec{k},t)$ and then their combination,
for the electric field operator we find
\begin{eqnarray}\label{c40}
&&\int d^3q \left[ \vec{k}\times \underline{\tilde{\mu}}(\vec{k},
\vec{q},s)\left(\vec{q}\times\underline{\vec{E}}^{ b}(\vec{q}
,s)\right)\right] -\frac{s^2}{c^2}\int d^3q \
\underline{\tilde{\varepsilon}}( \vec{k},
\vec{q},s)\underline{\vec{E}}^{b}(\vec{q}
,s)=\underline{\vec{J}}^{ b}_N(\vec{k},s)\nonumber\\
&&\int d^3q \left[ \vec{k}\times \underline{\tilde{\mu}}(\vec{k},
\vec{q},s)\left(\vec{q}\times\underline{\vec{E}}^{ f}(\vec{q}
,s)\right)\right] -\frac{s^2}{c^2}\int d^3q \
\underline{\tilde{\varepsilon}}( \vec{k},
\vec{q},s)\underline{\vec{E}}^{f}(\vec{q}
,s)=\underline{\vec{J}}^{ f}_N(\vec{k},s)\nonumber\\
\end{eqnarray}
where
\begin{eqnarray}\label{c41}
\underline{\tilde{\varepsilon}}( \vec{k},
\vec{q},s)&=&\delta(\vec{k}-\vec{q})I+\underline{\tilde{\chi}}^e(\vec{k},
\vec{q},s)\nonumber\\
\underline{\tilde{\mu}}( \vec{k},
\vec{q},s)&=&\delta(\vec{k}-\vec{q})I-\underline{\tilde{\chi}}^m(\vec{k},
\vec{q},s)
\end{eqnarray}
are respectively  the permitivity and permeability tensors of the
medium in Laplace language. The source terms in the right hand of
the  inhomogeneous Eqs.(\ref{c40})  are   the backward and forward
Laplace transformations of the noise current density and are
expressed in terms of initial conditions of the dynamical variables
of the total system at $t=0$ as
\begin{eqnarray}\label{c42}
\underline{\vec{J}}^{ b}_N(\vec{k},s)&=&\mu_0
s^2\underline{\vec{P}}^b_N(\vec{k},s)-\mu_0 s\
\imath\vec{k}\times\underline{M}^b_N(\vec{k},s)+\nonumber\\
&&\imath\vec{k}\times \int d^3q \ \underline{\tilde{\mu}}(\vec{k},
\vec{q},s)\
\underline{\vec{B}}(\vec{q},0)-\mu_0 s \underline{\vec{D}}(\vec{k},0)\nonumber\\
\underline{\vec{J}}^{ f}_N(\vec{k},s)&=&\mu_0 s^2\
\underline{\vec{P}}^f_N(\vec{k},s)+\mu_0 s\
\imath\vec{k}\times\underline{M}^f_N(\vec{k},s)\nonumber\\
&&-\imath\vec{k}\times \int d^3q \ \underline{\tilde{\mu}}(\vec{k},
\vec{q},s)\ \underline{\vec{B}}(\vec{q},0)-\mu_0 s
\underline{\vec{D}}(\vec{k},0)
\end{eqnarray}
Eqs. (\ref{c40}) can be rewritten in a compact form as
\begin{eqnarray}\label{c43}
\int d^3q \ \underline{\Lambda}(\vec{k},\vec{q},s)\ \cdot\
\underline{\vec{E}}^{b}(\vec{q},s)&=&\underline{\vec{J}}^{b}_N(\vec{k},s)\nonumber\\
\int d^3q \ \underline{\Lambda}(\vec{k},\vec{q},s)\ \cdot\
\underline{\vec{E}}^{f}(\vec{q},s)&=&\underline{\vec{J}}^{f}_N(\vec{k},s)
\end{eqnarray}
with
\begin{equation}\label{c44}
\underline {\Lambda}_{ij}(\vec{k},\vec{q},s)=
\varepsilon_{i\delta\gamma}\varepsilon_{\alpha\beta j}k^\delta
q^\beta
\underline{\tilde{\mu}}^{\gamma\alpha}(\vec{k},\vec{q},s)-\frac{s^2}{c^2}\underline{\tilde{\varepsilon}}_{ij}(\vec{k},\vec{q},s)
\end{equation}
 From (\ref{c43}) we see that the backward and
forward Laplace transformations of the electric field operator
satisfy an inhomogeneous equation with a source term. To solve such
equations it is common to  use the Green function method. The
suitable Green function associated to the Eq. (\ref{c43}) is defined
as the solution of an algebraic equation as follows
\begin{equation}\label{c45}
\int d^3q \ \underline{\Lambda}(\vec{k},\vec{q},s)\ \cdot\
\underline{G}(\vec{q}, \vec{p},s)= I \ \delta(\vec{k}-\vec{p})
\end{equation}
Accordingly  the solution of Eqs. (\ref{c43}) in terms of the Green
function defined in (\ref{c45}) can be written as
\begin{eqnarray}\label{c46}
\underline{\vec{E}}^{b}(\vec{k},s)&=& \int d^3p\
\underline{G}(\vec{k},\vec{p},s) \cdot\
\underline{\vec{J}}^b_N(\vec{p},s)\nonumber\\
\underline{\vec{E}}^{f}(\vec{k},s)&=& \int d^3p\
\underline{G}(\vec{k},\vec{p},s) \cdot\
\underline{\vec{J}}^f_N(\vec{p},s)
\end{eqnarray}
In the special case of a homogeneous medium for which the electric
and magnetic susceptibility tensors are translationally invariant,
that is when the tensors $ \chi^e(\vec{r},\vec{r'},t)$ and $
\chi^m(\vec{r},\vec{r'},t)$ are a function of the difference $
\vec{r}-\vec{r'}$ and so is then, the permitivity and permeability
tensors $ \varepsilon (\vec{r},\vec{r'},t)$ and $
\mu(\vec{r},\vec{r'},t) $, we deduce
\begin{eqnarray}\label{c47}
 \underline{\tilde{\varepsilon}}(\vec{k},\vec{q} ,s)&=&
\left(I+\underline{\tilde{\chi}}^e(\vec{k},s)\right)\delta(\vec{k}-\vec{q})\nonumber\\
\underline{\tilde{\mu}}(\vec{k},\vec{q}
,s)&=&\left(I-\underline{\tilde{\chi}}^m(\vec{k},s)\right)\delta(\vec{k}-\vec{q})
\end{eqnarray}
and therefore from the Eqs. (\ref{c44}) and (\ref{c45})  we find
\begin{eqnarray}\label{c48}
&&\underline{G}(\vec{k},\vec{p},s)=
\underline{T}^{-1}(\vec{k},s)\delta(\vec{k}-\vec{p})\nonumber\\
&&\underline{T}_{ij}=\left[\varepsilon_{i\delta\gamma}\varepsilon_{\alpha\beta
j}k_\delta k_\beta \left(
\delta_{\gamma\alpha}-\underline{\tilde{\chi}}^m_{\gamma\alpha}(\vec{k},s)\right)-
\frac{s^2}{c^2}\left(\delta_{ij}+\underline{\tilde{\chi}}^e_{ij}(\vec{k},s)\right)\right]\nonumber\\
\end{eqnarray}
where it should be summed on the indices $ \alpha,\beta , \gamma
,\delta $ in the first term in the bracket.
 For a general medium it may be impossible  to solve Eq.(\ref{c45})
 exactly to obtain the Green function $ \underline{G} $. However one can use a suitable iteration method
 and find the Green function up to an arbitrary accuracy which may
 be useful for media with considerably weak polarizability. Let us
 write the tensor $ \underline{\Lambda}(\vec{k},\vec{q},s)$ as the sum of two
 parts
\begin{equation}\label{c49}
\underline{\Lambda}(\vec{k},\vec{q},s)=\underline{\Lambda}^{(0)}(\vec{k},\vec{q},s)+\underline{\Lambda}^{(1)}(\vec{k},\vec{q},s)
\end{equation}
where
\begin{equation}\label{c50}
\underline{\Lambda}^{(0)}_{ij}(\vec{k},\vec{q},s)=\left(
k_ik_j-\delta_{ij}k^2-\frac{s^2}{c^2}\delta_{ij}\right)\delta(\vec{k}-\vec{q})\equiv
L(\vec{k},s)\delta(\vec{k}-\vec{q})
\end{equation}
is the tensor $\Lambda $ in free space, that is when there is no
medium  and
\begin{equation}\label{c51}
\underline{\Lambda}^{(1)}_{ij}(\vec{k},\vec{q},s)=-\varepsilon_{i\delta\gamma}\varepsilon_{\alpha\beta
j}k_\delta q_\beta
\underline{\tilde{\chi}}^m_{\gamma\alpha}(\vec{k},\vec{q},s)-\frac{s^2}{c^2}\underline{\tilde{\chi}}^e_{ij}(\vec{k},\vec{q},s)
\end{equation}
is the part due to  the presence of the medium, that is  the effect
of the medium in the total tensor $\Lambda $. Then, following  the
Born approximation method previously used  in the scattering theory
\cite{23}, the Green function $\underline{G}$ can be expressed as a
series as follows
\begin{equation}\label{c52}
\underline{G}(\vec{k},\vec{p},s)=\underline{G}^{(0)}(\vec{k},\vec{p},s)+\underline{G}^{(1)}(\vec{k},\vec{p},s)
+ \underline{G}^{(2)}(\vec{k},\vec{p},s)+\ \cdots\
\end{equation}
where
\begin{equation}\label{c53}
\underline{G}^{(0)}(\vec{k},\vec{p},s)=L^{-1}(\vec{k},s)\
\delta(\vec{k}-\vec{p})
\end{equation}
is the Green function for the free space  and $
\underline{G}^{(n)}(\vec{k},\vec{p},s)\hspace{00.50
cm}n=1,2,3\cdots $ are found  from the following recurrence
relation
\begin{eqnarray}\label{c54}
&&\int d^3q \ \underline{\Lambda}^{(0)}(\vec{k},\vec{q},s)\ \cdot\
\underline{G}^{(n)}(\vec{q}, \vec{p},s)= -\int d^3q \
\underline{\Lambda}^{(1)}(\vec{k},\vec{q},s)\ \cdot\
\underline{G}^{(n-1)}(\vec{q}, \vec{p},s)\nonumber\\
&&\Longrightarrow \underline{G}^{(n)}(\vec{k},
\vec{p},s)=-L^{(-1)}(\vec{k},s)\int d^3q\
\underline{\Lambda}^{(1)}(\vec{k},\vec{q},s)\ \cdot\
\underline{G}^{(n-1)}(\vec{q}, \vec{p},s)
\end{eqnarray}
where we have used  Eq.(\ref{c50}). Using the series (\ref{c52}) and
the recurrence relation (\ref{c54}) one can obtain the Green
function $\underline{G}$ up to an arbitrary accuracy. In n'th order
approximation it can be neglected  from the terms
$\underline{G}^{(j)}\hspace{00.50 cm}j\geq n+1$ in the series
(\ref{c52}) which may be useful for media with considerably weak
polarizability.\\
Now, having the Green function $\underline{G}$, the time dependence
of electric field operator for negative and positive times  can be
obtained by inverse Laplace transformations of $
\underline{\vec{E}}^b$ and $ \underline{\vec{E}}^f$ , respectively.
Carrying out the inverse Laplace transformation of Eq. (\ref{c46})
for $ t<0$ we deduce
\begin{equation}\label{c55}
\underline{\vec{E}}(\vec{k},t)=\frac{1}{2\pi}\int_{-\infty}^{+\infty}
d\omega\ e^{-\imath\omega t}\int d^3p\ \underline{G}(\vec{k},\
\vec{p},\ \imath\omega+0^+) \cdot\
\underline{\vec{J}}^b_N(\vec{p},\ \imath\omega+0^+)
\end{equation}
where $0^+$ is an arbitrarily  small positive number. For $t>0$ the
inverse Laplace transformation of $ \underline{\vec{E}}^f$ is
\begin{equation}\label{c56}
\underline{\vec{E}}(\vec{k},t)=\frac{1}{2\pi}\int_{-\infty}^{+\infty}
d\omega\ e^{-\imath\omega t}\int d^3p\ \underline{G}(\vec{k},\
\vec{p},\ -\imath\omega+0^+) \cdot\
\underline{\vec{J}}^f_N(\vec{p},\ -\imath\omega+0^+)
\end{equation}
Because the Laplace transformations of the susceptibility tensors
are analytic functions respect to the variable  $s$ in any point of
the half-plane $ Re[s]\geq 0$ then, clearly  the $\omega$-dependence
integrand in Eq. (\ref{c55}) also is analytic in the  half-plane $Im
[\omega] \leq 0 $. Accordingly, the integral over $\omega$ in
(\ref{c55}) vanishes for the positive times . Likewise the integral
in (\ref{c56}) is zero for $t<0$. Therefore we can combine the two
expressions (\ref{c55}) and  (\ref{c56}) into a single one and write
the full time-dependence of electric field for all times as
\begin{eqnarray}\label{c57}
\underline{\vec{E}}(\vec{k},t)&=&\frac{1}{2\pi}\int_{-\infty}^{+\infty}
d\omega\ e^{-\imath\omega t}\int d^3p\
\left[\underline{G}(\vec{k},\ \vec{p},\ \imath\omega+0^+) \cdot\
\underline{\vec{J}}^b_N(\vec{p},\ \imath\omega+0^+)\right.\nonumber\\
&+& \left.\underline{G}(\vec{k},\ \vec{p},\ -\imath\omega+0^+)
\cdot\ \underline{\vec{J}}^f_N(\vec{p},\ -\imath\omega+0^+)\right]
\end{eqnarray}
It is remarkable that using the asymptotic behavior of the Green
function $\underline{G}$ and the susceptibility tensors for large
$|\omega|$ it can be verified straightforwardly,  the integral in
(\ref{c57}) reduces to $\underline{\vec{E}}(\vec{k},0)$ at $t=0$ and
hence this equation give us correctly the time dependence of the
electric field for all times. Finally having the electric field
operator, we can obtain the magnetic field using $
\imath\vec{k}\times\underline{\vec{E}}(\vec{k},t)=\underline{\dot{\vec{B}}}(\vec{k},t)$
and then the polarization densities employing  the constitutive
Eqs.(\ref{c31}) and (\ref{c36}).
\section{The magnetodielectric  media with only one response equation}
In the previous sections we have assumed the magnetodielectric
medium under consideration is one for  which the distinction between
polarization and magnetization physically is possible and the total
 current induced in medium can be separated into electric and magnetic parts.
It is remarkable that the use of polarization and magnetization ,
separately, to describe the response of a medium to electromagnetic
field is useful for media which are  not spatially dispersive  or
which have a particular form of spatial dispersion \cite{24}. There
are in general  media for which it is not clear how the separation
of the total current induced in medium is to be made into electric
and magnetic parts. The disturbances, caused by the electromagnetic
field, in such media  are described completely in terms of only a
response equation relating the total current  induced in medium to
the electric field \cite{24}. The quantization method outlined in
the previous sections can cover such cases applying the Lagrangian
\begin{eqnarray}\label{c58}
 L(t)&=& \int _0^\infty d\omega \int d^3r \left[ \frac{1}{2}
 \dot{\vec{X}}_\omega \cdot  \dot{\vec{X}}_\omega
 -\frac{1}{2}\omega^2
 \vec{X}_\omega \cdot  \vec{X}_\omega\right]\nonumber\\
&+&\int d^3r \left[ \frac{1}{2}\varepsilon _0
\vec{E}^2-\frac{\vec{B}^2}{2\mu_0}\right]\nonumber\\
&+& \int _0^\infty d \omega \int d^3 r \int d^3 r' f_{ij}( \omega ,
\vec{r} , \vec{r'}) E^i(\vec{r},t) \vec{X}_\omega ^j(\vec{r'} ,t)
\end{eqnarray}
which is the same as the Lagrangian (\ref{c1})-(\ref{c4}) with the
exceptional that, now the medium is modeled   by a single set of
three dimensional harmonic oscillators $ \vec{X}_\omega$. In this
case the Euler- Lagrange equation (\ref{c11}) takes the form
\begin{eqnarray}\label{c59}
&& \frac{d}{d t}\left(\frac{\delta\underline{L}}{\delta
\left(\underline{\dot{A}}^*_i(\vec{k},t)\right)}\right)-\frac{\delta\underline{L}}{\delta\left(\underline{A}^*_i(\vec{k},t)\right)}=0
\hspace{2.00 cm} i=1,2,3\nonumber\\
\Longrightarrow&&\mu_0\varepsilon_0\underline{\ddot{\vec{A}}}(\vec{k},t)+\mu_0\varepsilon_0
\ \imath\vec{k}\
\underline{\dot{\varphi}}(\vec{k},t)-\vec{k}\times\left(\vec{k}\times\underline{\vec{A}}(\vec{k},t)\right)=
\mu_0\ \underline{\dot{\vec{P}}}(\vec{k},t)\nonumber\\
&&
\end{eqnarray}
where now
\begin{equation}\label{c60}
\underline{\vec{P}}(\vec{k},t)=\int_0^\infty d\omega\int d^3p \
\underline{f}( \omega , \vec{k},\vec{p})\cdot\
\underline{X}_\omega(\vec{p},t)
\end{equation}
can no longer be interpreted as the electric polarization, but is
related to the total current induced in the medium by
$\underline{j}(\vec{k},t)=\underline{\dot{\vec{P}}}(\vec{k},t)$.
Following the canonical quantization similar to the previous
sections we obtain the only response equation of the medium as
\begin{equation}\label{c61}
\underline{\vec{P}}(\vec{k},t)=\underline{\vec{P}}_N(\vec{k},t)+\varepsilon_0\int_0^{|t|}
dt' \int d^3q\ \underline{\chi}(\vec{k},\vec{q},
|t|-t')\cdot\underline{\vec{E}}(\vec{q},\pm t')
\end{equation}
instead of the two constitutive equations (\ref{c31}) and
(\ref{c38}), where $ \underline{\vec{P}}_N(\vec{k},t)$ is given by
(\ref{c35}) and the response tensor
\begin{equation}\label{c62}
\underline{\chi}(\vec{k},\vec{q}, t)=\left\{
\begin{array}{cc}
  \frac{1}{\varepsilon_0}\int_0^\infty d\omega \frac{\sin \omega t}{\omega}\int d^3p \ \underline{f}(\omega , \vec{k} , \vec{p}) \cdot
  \underline{f}^\dag(\omega , \vec{q} , \vec{p}) & t> 0 \\
  \\
  0 & t\leq 0
\end{array}\right.
\end{equation}
is related to the conductivity tensor of the medium by $
\frac{\partial\underline{\chi}}{\partial t}=\underline{\sigma}$.
This is one of the two  alternative descriptions of the response of
a magnetodielectric medium  for which the distinction between
polarization and magnetization is not possible \cite{24}. Finally
following the same calculations as before sections one can obtain
the space -time dependence of electromagnetic field operators in
terms of the response tensor $\underline{\chi}$ and the dynamical
variables $ \underline{\vec{X}}_\omega$ at $t=0$. The calculations
in this case are precisely the same as before sections  except that
the quantities $ \underline{\vec{Y}}_\omega ,
\underline{\vec{\Pi}}_\omega , \underline{\vec{M}},
\underline{\chi}^m , \underline{g}$ are identically zero.
\section{Summary and conclusion}
By modeling an anisotropic polarizable and magnetizable medium with
two continuous collections  of three dimensional vector fields, we
have presented  a fully canonical quantization for electromagnetic
field in the presence of such a medium. In this model the
polarization fields of the medium did   not enter directly in the
 Lagrangian of the total system as a part of the degrees of
freedom of the medium, but the space-dependent oscillators modeling
the medium solely was sufficient to describe both polarizability and
the absorption of the medium. The interaction of the medium with the
electromagnetic field explicitly was given and both the Maxwell's
laws and the constitutive equations of the medium were obtained as
the consequences of the Heisenberg equations of total system. The
electric and magnetic polarization fields of the medium could be
deduced    naturally in terms of the dynamical fields modeling the
medium. Some space dependent real valued tensor
 coupling the medium with electromagnetic
field were introduced. The coupling tensor had  an important role in
this quantization scheme, so that the electric and magnetic
susceptibility tensors of the medium were determined in terms of the
coupling tensor. Also the noise polarizations were expressed in
terms of the coupling tensor and the dynamical variables describing
the degrees of freedom of the medium at $t=0$. In the free space and
in the case of a non-absorbing medium the coupling tensor and the
noise polarizations tend to zero and this quantization method is
reduced to the usual quantization in these limiting cases. As a
tool, we have used the reciprocal space to quantize the total
system. It was shown the temporal backward and forward Laplace
transforms of the electric field  obey some algebraic equations with
source terms that could be determined in terms of the dynamical
variables of the system at $t=0$. By introducing the Green function
of these algebraic equations we were able to determine the Green
function up to an arbitrary accuracy using a perturbation  method
which may be useful in some real cases with weak polarizability. The
time-dependence of electric field was derived for both negative and
positive times. Finally the model was modified  to the case of a
 medium  for which the distinction between
the polarization and magnetization is not possible.

\end{document}